\documentclass{itg} % Change this for final version!

\usepackage{amsmath,amscd,amssymb}
\usepackage{fancyhdr,graphics,pifont}
\usepackage{floatflt,subfigure,epsfig,moreverb,multicol,lscape}
\usepackage{supertabular,tabularx,multirow,bar}
\usepackage{psfrag,afterpage,curves,epic, bbm, stmaryrd}
\usepackage[english]{babel}

\newcommand{\supp}{\operatorname{supp}}

\newcommand{\matr}[1]{\mathbf{#1}}
\newcommand{\vect}[1]{\mathbf{#1}}
\newcommand{\code}[1]{\mathcal{#1}}

\newcommand{\set}[1]{\mathcal{#1}}
\newcommand{\graph}[1]{\mathsf{#1}}

\newcommand{\GF}[1]{\mathbb{F}_{#1}}
\newcommand{\R}{\mathbb{R}}
\newcommand{\Rp}{\mathbb{R}_{+}}
\newcommand{\Rpp}{\mathbb{R}_{++}}

\newcommand{\defeq}{\triangleq}

\newtheorem{Definition}{Definition}

\newtheorem{Lemma}[Definition]{Lemma}

\newtheorem{Proposition}[Definition]{Proposition}

\newtheorem{Conjecture}[Definition]{Conjecture}

\newenvironment{Proof}%
  {\noindent \emph{Proof:}}{\hfill$\square$}

\newcommand{\eproposition}{\hfill$\square$}

\newcommand{\edefinition}{\mbox{ }\hfill$\square$}
\newcommand{\econjecture}{\hfill$\square$}

\newcommand{\convhull}{\operatorname{conv}}

\newcommand{\va}{\vect{a}}
\newcommand{\tva}{\tilde{\vect{a}}}
\newcommand{\vb}{\vect{b}}
\newcommand{\tvb}{\tilde{\vect{b}}}

\newcommand{\vd}{\vect{d}}

\newcommand{\vg}{\vect{g}}

\newcommand{\vu}{\vect{u}}

\newcommand{\tvu}{\tilde{\vect{u}}}

\newcommand{\vv}{\vect{v}}
\newcommand{\tvv}{\tilde{\vect{v}}}
\newcommand{\uvv}{\vect{\breve{v}}}

\newcommand{\vX}{\vect{X}}
\newcommand{\vx}{\vect{x}}

\newcommand{\vY}{\vect{Y}}
\newcommand{\vy}{\vect{y}}

\newcommand{\vlambda}{\boldsymbol{\lambda}}

\newcommand{\vxi}{\boldsymbol{\xi}}

\newcommand{\tr}{\mathsf{T}}

\newcommand{\e}{\mathrm{e}}

\newcommand{\optprog}[2]
{%
  \noindent\mbox{}\\[0cm]
  \noindent\fbox{%
  \begin{minipage}{0.955\linewidth}
    \mbox{}\\[-0.5cm]
    #1\\[#2]
  \end{minipage}
  }
  \noindent\mbox{}\\[-0.2cm]
}

\begin{document}

\title{Towards Low-Complexity Linear-Programming Decoding}

\author{Pascal O.~Vontobel$^1$ and Ralf Koetter$^2$ \\[1mm]
        \mbox{}$^1$ Dept.~of EECS,
                    MIT,
                    Cambridge, MA 02139, USA,
                    \texttt{pascal.vontobel@ieee.org}. \\
        \mbox{}$^2$ CSL and Dept.~of ECE,
                    University of Illinois,
                    Urbana, IL 61801, USA,
                    \texttt{koetter@uiuc.edu}.
}

\thanks{P.O.V.'s research was supported by NSF Grants TF 05-14801,
  ATM-0296033, DOE SciDAC, and ONR Grant N00014-00-1-0966. The research for
  this paper was partly done while being at the Dept.~of ECE, University of
  Wisconsin-Madison, Madison, WI 53706, USA. R.K.'s research was
  supported by NSF Grants CCR 99-84515, CCR 01-05719, and TF 05-14869.}

\abstract{We consider linear-programming (LP) decoding of low-density
  parity-check (LDPC) codes. While it is clear that one can use any
  general-purpose LP solver to solve the LP that appears in the
  decoding problem, we argue in this paper that the LP at hand is
  equipped with a lot of structure that one should take advantage of.
  Towards this goal, we study the dual LP and show how
  coordinate-ascent methods lead to very simple update rules that are
  tightly connected to the min-sum algorithm. Moreover, replacing
  minima in the formula of the dual LP with soft-minima one obtains
  update rules that are tightly connected to the sum-product
  algorithm. This shows that LP solvers with complexity similar to the
  min-sum algorithm and the sum-product algorithm are feasible.
  Finally, we also discuss some sub-gradient-based methods.}

\maketitle

\section{Introduction}
\label{sec:introduction:1}

Linear-programming (LP) decoding~\cite{Feldman:03:1,
Feldman:Wainwright:Karger:05:1} has recently emerged as an interesting option
for decoding low-density parity-check (LDPC) codes. Indeed, the observations
in ~\cite{Koetter:Vontobel:03:1, Vontobel:Koetter:05:1:subm,
Vontobel:Koetter:04:2} suggest that the LP decoding performance is very close
to the message-passing iterative (MPI) decoding performance. Of course, one
can use any general-purpose LP solver to solve the LP that appears in LP
decoding, however in this paper we will argue that one should take
advantage of the special structure of the LP at hand in order to formulate
efficient algorithms that provably find the optimum of the LP.

Feldman et al.~\cite{Feldman:Karger:Wainwright:02:1} briefly mention the use
of sub-gradient methods for solving the LP of an early version of the LP
decoder (namely for turbo-like codes). Moreover, Yang et
al.~\cite{Yang:Wang:Feldman:05:1} present a variety of interesting approaches
to solve the LP where they use some of the special features of the LP at
hand. However, we belive that one can take much more advantage of the
structure that is present: this paper shows some results in that direction.

So far, MPI decoding has been successfully used in applications where
block error rates on the order of $10^{-5}$ are needed because for
these block error rates the performance of MPI decoding can be
guaranteed by simulation results. However, for applications like
magnetic recording, where one desires to have block error rates on the
order of $10^{-15}$ and less, it is very difficult to guarantee that
MPI decoding achieves such low block error rates for a given
signal-to-noise ratio. The problem is that simulations are too
time-consuming and that the known analytical results are not strong
enough. Our hope and main motivation for the present work is that
efficient LP decoders, together with analytical results on LP decoding
(see e.g.~\cite{Vontobel:Koetter:04:1, Chaichanavong:Siegel:05:1,
  Vontobel:Smarandache:05:1}), can show that efficient decoders exist
for which low block error rates can be guaranteed for a certain
signal-to-noise ratio.

This paper is structured as follows. We start off by introducing in
Sec.~\ref{sec:primal:linear:program:1} the primal LP that appears in LP
decoding. In Sec.~\ref{sec:dual:linear:program:1} we formulate the dual LP and
in Secs.~\ref{sec:softened:dual:linear:program:1}
and~\ref{sec:primal:softened:dual:linear:program:1} we consider a ``softened''
version of this dual LP. Then, in Secs.~\ref{sec:decoding:algorithm:1}
and~\ref{sec:decoding:algorithm:2} we propose some efficient decoding
algorithms and in Sec.~\ref{sec:simulation:results:1} we show some simulation
results. Finally, in Sec.~\ref{sec:conclusions:1} we offer some conclusions
and in the appendix we present the proofs and some additional material.

Before going to the main part of the paper, let us fix some notation. We let
$\R$, $\Rp$, and $\Rpp$ be the set of real numbers, the set of non-negative
real numbers, and the set of positive real numbers, respectively. Moreover, we
will use the canonical embedding of the set $\GF{2} = \{ 0, 1 \}$ into
$\R$. The convex hull of a set $\set{A} \subseteq \R^n$ is denoted by
$\convhull(\set{A})$. If $\set{A}$ is a subset of $\GF{2}^n$ then
$\convhull(\set{A})$ denotes the convex hull of the set $\set{A}$ after
$\set{A}$ has been canonically embedded in $\R^n$. The $i$-th component of a
vector $\vx$ will be called $[\vx]_i$ and the element in the $j$-th row and
$i$-th column of a matrix $\matr{A}$ will be called $[\matr{A}]_{j,i}$.

Moreover, we will use Iverson's convention, i.e. for a statement $A$ we have
$[A] = 1$ if $A$ is true and $[A] = 0$ otherwise. From this we also derive the
notation $\big\llbracket A \big\rrbracket \defeq -\log [A]$,
i.e.~$\big\llbracket A \big\rrbracket = 0$ if $A$ is true and $\big\llbracket
A \big\rrbracket = +\infty$ otherwise. Let $\set{A}$ and $\set{X}$ be some
arbitrary sets fulfilling $\set{A} \subseteq \set{X}$. A function like
$\set{X} \to \Rp: \vx \mapsto [\vx \in \set{A}]$ is called an indicator
function for the set $\set{A}$, whereas a function like $\set{X} \to \Rp: \vx
\mapsto \big\llbracket \vx \in \set{A} \big\rrbracket$ is called a neglog
indicator function for the set $\set{A}$. Of course, this second function can
also be considered as a cost or penalty function.

Throughout the paper, we will consider a binary linear code $\code{C}$ that is
defined by a parity-check matrix $\matr{H}$ of size $m$ by $n$. Based on
$\matr{H}$, we define the sets $\set{I} \defeq \set{I}(\matr{H}) \defeq \{ 1,
\ldots, n \}$, $\set{J} \defeq \set{J}(\matr{H}) \defeq \{ 1, \ldots, m \}$,
$\set{I}_j \defeq \set{I}_j(\matr{H}) \defeq \{ i \in \set{I} \ | \
[\matr{H}]_{j,i} = 1 \}$ for each $j \in \set{J}$, $\set{J}_i \defeq
\set{J}_i(\matr{H}) \defeq \{ j \in \set{J} \ | \ [\matr{H}]_{j,i} = 1 \}$ for
each $i \in \set{I}$, and $\set{E} \defeq \set{E}(\matr{H}) \defeq \set\{
(i,j) \in \set{I} \times \set{J} \ | \ i \in \set{I}, j \in \set{J}_i \} =
\set\{ (i,j) \in \set{I} \times \set{J} \ | \ j \in \set{J}, i \in \set{I}_j
\}$.  Moreover, for each $j \in \set{J}$ we define the codes $\code{C}_j
\defeq \code{C}_j(\matr{H}) \defeq \{ \vx \in \GF{2}^n \ | \ \vect{h}_j
\vx^\tr = 0 \text{ (mod $2$)} \}$, where $\vect{h}_j$ is the $j$-th row of
$\matr{H}$. Note that the code $\code{C}_j$ is a code of length $n$ where all
positions not in $\set{I}_j$ are unconstrained.

We will express the linear programs in this paper in the framework of
Forney-style factor graphs (FFG)~\cite{Kschischang:Frey:Loeliger:01,
Forney:01:1, Loeliger:04:1}, sometimes also called normal graphs. For
completeness we state their formal definition. An FFG is a graph
$\graph{G}(V,E)$ with vertex set $V$ and edge set $E$. To each edge $e$ in the
graph we associate a variable $x_e$ defined over a suitably chosen alphabet
$\set{X}_e$. Let $v$ be a node in the FFG and let $E_v$ be the set of edges
incident to $v$. Any node $v$ in the graph is associated with a function $f_v$
with domain $\set{X}_{e_1} \times \set{X}_{e_2} \times \cdots \times
\set{X}_{e_\ell}$ where $\{e_1,e_2,\ldots,e_\ell\}=E_v$. The co-domain of
$f_v$ is typically $\R$ or $\R_+$.

FFGs typically come in two flavors, either representing the factorization of a
function into a product of terms or a decomposition of an additive cost
function. In our case we will exclusively deal with the latter case. The
global function $g(x_{e_1},x_{e_2},\ldots,x_{e_{|E|}})$ represented by an FFG
is then given by the sum $g(x_{e_1},x_{e_2},\ldots,x_{e_{|E|}}) \defeq
\sum_{v\in V} f_v$.

\section{The Primal Linear Program}
\label{sec:primal:linear:program:1}

\begin{figure}[t]
  \begin{center}
    \epsfig{file=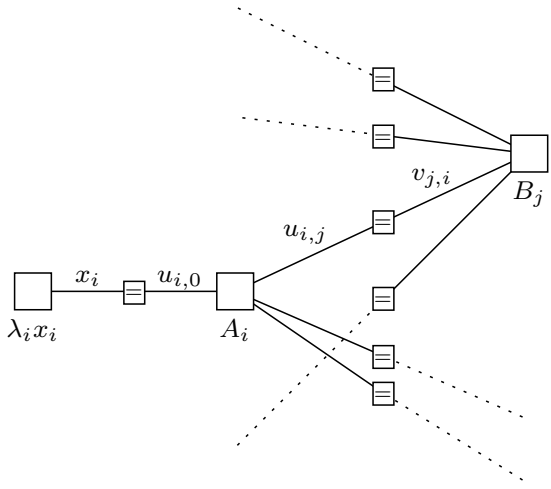, height=0.80\linewidth}
  \end{center}
  \caption{Representative part of the FFG for the augmented cost function
  in~\eqref{eq:primal:lp:augmented:cost:function:1}. (Note that this FFG has
  an additively written global function.)}
  \label{fig:ffg:ldpc:code:binary:1:1}
\end{figure}

The code $\code{C}$ is used for data transmission over a binary-input
memoryless channel with channel law $P_{\vect{Y} | \vect{X}}(\vect{y}
| \vect{x}) = \prod_{i \in \set{I}} P_{Y | X}(y_i | x_i)$. Upon
observing $\vect{Y} = \vect{y}$, the maximum-likelihood decoding (MLD)
rule decides for $\hat \vx(\vy) = \arg \max_{\vx \in \code{C}}
P_{\vect{Y} | \vect{X}}(\vect{y} | \vect{x})$. This can also be
written as

\optprog
{
\begin{alignat*}{2}
  \textbf{MLD1:}
  \quad
  &
  \text{maximize} \quad
  &&
    P_{\vY |\vX}(\vy | \vx) \\
  &
  \text{subject to } \quad
  &&
   \vx \in \code{C}.
\end{alignat*}
}{-0.8cm}

\noindent It is clear that instead of $P_{\vY |\vX}(\vy | \vx)$ we can also
maximize $\log P_{\vY |\vX}(\vy | \vx) = \sum_{i \in \set{I}} \log P_{Y |
X}(y_i | x_i)$. \noindent Introducing $\lambda_i \defeq \lambda_i(y_i) \defeq
\log \bigl( \frac{P_{Y | X}(y_i|0)}{P_{Y | X}(y_i | 1)} \bigr)$, $i \in
\set{I}$, and noting that $\log P_{Y | X}(y_i | x_i) = - \lambda_i x_i + \log
P_{Y | X}(y_i | 0), \textbf{MLD1}$ can then be rewritten to read

\optprog
{
\begin{alignat*}{2}
  \textbf{MLD2:}
  \quad
  &
  \text{minimize} \quad
  &&
    \sum_{i \in \set{I}}
      \lambda_i x_i \\
  &
  \text{subject to } \quad
  &&
   \vx \in \code{C}.
\end{alignat*}
}{-0.8cm}

\begin{figure}[t]
  \begin{center}
    \epsfig{file=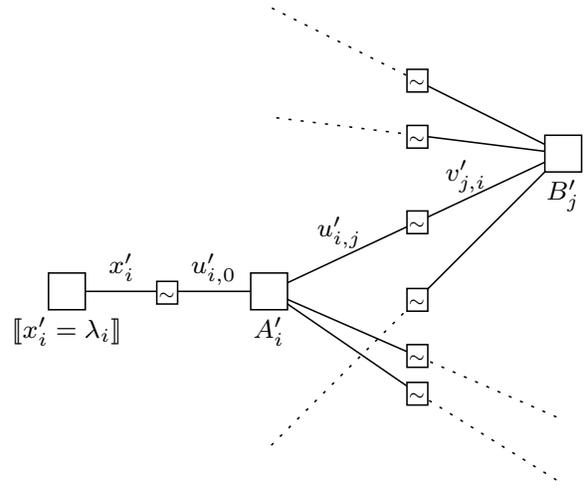, height=0.80\linewidth}
  \end{center}
  \caption{Representative part of the FFG for the augmented cost function
    in~\eqref{eq:dual:lp:augmented:cost:function:1}. Function nodes with a
    tilde sign in them mean the following: if such a function node is
    connected to edges $u$ and $v$ then the function value is $- \llbracket u
    = -v \rrbracket$. (Note that this FFG has an additively written global
    function.)}
  \label{fig:dual:ffg:ldpc:code:binary:1:1}
\end{figure}

\noindent Because the cost function is linear, and a linear function attains
its minimum at the extremal points of a convex set, this is essentially
equivalent to

\optprog
{
\begin{alignat*}{2}
  \textbf{MLD3:}
  \quad
  &
  \text{minimize} \quad
  &&
    \sum_{i \in \set{I}}
      \lambda_i x_i \\
  &
  \text{subject to } \quad
  &&
   \vx \in \convhull(\code{C}).
\end{alignat*}
}{-0.8cm}

\noindent Although this is a linear program, it can usually not be solved
efficiently because its description complexity is usually exponential in the
block length of the code.

However, one might try to solve a relaxation of \textbf{MLD3}. Noting that
$\convhull(\code{C}) \subseteq \convhull(\code{C}_1) \cap \cdots \cap
\convhull(\code{C}_m)$ (which follows from the fact that $\code{C} =
\code{C}_1 \cap \cdots \cap \code{C}_m$), Feldman, Wainwright, and
Karger~\cite{Feldman:03:1, Feldman:Wainwright:Karger:05:1} defined the
(primal) linear programming decoder (PLPD) to be given by the solution of the
linear program

\optprog
{
\begin{alignat*}{2}
  \textbf{PLPD1:}
  \quad
  &
  \text{minimize} \quad
  &&
    \sum_{i \in \set{I}}
      \lambda_i x_i \\
  &
  \text{subject to } \quad
  &&
   \vx \in \convhull(\code{C}_j)
     \ (j \in \set{J}).
\end{alignat*}
}{-0.8cm}

\noindent The inequalities that are implied by the expression $\vx \in
\convhull(\code{C}_j)$ can be found in~\cite{Feldman:03:1,
Feldman:Wainwright:Karger:05:1, Koetter:Vontobel:03:1,
Vontobel:Koetter:05:1:subm}. Although \textbf{PLPD1} is usually suboptimal
compared to MLD, it is especially attractive for LDPC codes for two reasons:
firstly, for these codes the description complexities of
$\convhull(\code{C}_j)$, $j \in \set{J}$, turn out to be
low~\cite{Feldman:Wainwright:Karger:05:1, Vontobel:Koetter:05:1:subm} and,
secondly, the relaxation is relatively benign only if the weight of the parity
checks is low.  There are many ways of reformulating this \textbf{PLPD1} rule
by introducing auxiliary variables: one way that we found particularly useful
is shown as \textbf{PLPD2} below. The reason for its usefulness is that there
is a one-to-one correspondence between parts of the program and the FFG shown
in Fig.~\ref{fig:ffg:ldpc:code:binary:1:1}, as we will discuss later
on. Indeed, while the notation may seem heavy at first glance, it precisely
reflects the structure of the constraints that are summarily folded into the
seemingly simpler constraint $\vx \in \convhull(\code{C}_j) \ (j \in \set{J})$
of \textbf{PLPD1}.

\optprog
{
\begin{alignat*}{3}
  &
  \textbf{PLPD2:} \\
  &
  \text{min.} \quad
  &
    \sum_{i \in \set{I}}
      \lambda_i x_i \\
  &
  \text{subj.~to } \quad
  &
  x_i
    &= u_{i,0}
       \quad &&(i \in \set{I}), \\
  &&
  u_{i,j}
    &= v_{j,i}
       \quad &&((i,j) \in \set{E}), \\
  &&
  \sum_{\va_i \in \code{A}_i}
    \alpha_{i, \va_i} \va_i
    &= \vu_i
       \quad &&(i \in \set{I}), \\
  &&
  \sum_{\vb_j \in \code{B}_j}
    \beta_{j, \vb_j} \vb_j
    &= \vv_j
       \quad &&(j \in \set{J}), \\
  &&
  \alpha_{i, \va_i}
    &\geq 0
       \quad &&(i \in \set{I}, \va_i \in \code{A}_i), \\
  &&
  \beta_{j, \vb_j}
    &\geq 0
       \quad &&(j \in \set{J}, \vb_j \in \code{B}_j), \\
  &&
  \sum_{\va_i \in \code{A}_i}
    \alpha_{i, \va_i}
    &= 1
       \quad &&(i \in \set{I}), \\
  &&
  \sum_{\vb_j \in \code{B}_j}
    \beta_{j, \vb_j}
    &= 1
       \quad &&(j \in \set{J}).
\end{alignat*}
}{-0.6cm}

\noindent Here we used the following codes, variables and vectors. The code
$\code{A}_i \subseteq \{ 0, 1 \}^{|\{ 0 \} \cup \set{J}_i|}$, $i \in \set{I}$,
is the set containing the all-zeros vector and the all-ones vector of length
$|\set{J}_i| + 1$, and $\code{B}_j \subseteq \{ 0, 1 \}^{|\set{I}_j|}$, $j \in
\set{J}$, is the code $\code{C}_j$ shortened at the positions $\set{I}
\setminus \set{I}_j$.\footnote{For the codes $\code{C}$ under consideration
this means that $\code{B}_j$ contains all vectors of length $|\set{I}_j|$ of
even parity.} For $i \in \set{I}$ we will also use the vectors $\vu_i$ where
the entries are indexed by $\{ 0 \} \cup \set{J}_i$ and denoted by $u_{i,j}
\defeq [\vu_i]_j$, and for $j \in \set{J}$ we will use the vectors $\vv_j$
where the entries are indexed by $\set{I}_j$ and denoted by $v_{j,i} \defeq
[\vv_j]_i$. Later on, we will use a similar notation for the entries of
$\va_i$ and $\vb_j$, i.e.~we will use $a_{i,j} \defeq [\va_i]_j$ and $b_{j,i}
\defeq [\vb_j]_i$, respectively.

The above optimization problem is elegantly represented by the FFG shown in
Fig.~\ref{fig:ffg:ldpc:code:binary:1:1}. In order to express the LP itself in
an FFG we have to express the constraints as additive cost terms.  This is
easily accomplished by assigning the cost $+\infty$ to any configuration of
variables that does not satisfy the LP constraints.  The above minimization
problem is then equivalent to the (unconstrained) minimization of the
augmented cost function
\begin{align}
  \sum_{i \in \set{I}}
    \lambda_i x_i
  &
  +
  \sum_{i \in \set{I}}
    \big\llbracket
      x_i = u_{i,0}
    \big\rrbracket
  +
  \sum_{(i,j) \in \set{E}}
    \big\llbracket
      u_{i,j} = v_{j,i}
    \big\rrbracket
      \nonumber \\
  &
  +
  \sum_{i \in \set{I}}
    A_i(\vu_i)
  +
  \sum_{j \in \set{J}}
    B_j(\vv_j),
      \label{eq:primal:lp:augmented:cost:function:1}
\end{align}
where for all $i \in \set{I}$ and all $j \in \set{J}$, respectively, we
introduced
\begin{align*}
  A_i(\vu_i)
    &\defeq
           \left\llbracket
             \sum_{\va_i \in \code{A}_i}
               \alpha_{i, \va_i} \va_i
               = \vu_i
           \right\rrbracket \\
    &\quad\quad
           +
           \sum_{\va_i \in \code{A}_i}
             \big\llbracket
               \alpha_{i, \va_i}
                 \geq 0
             \big\rrbracket
           +
           \left\llbracket
             \sum_{\va_i \in \code{A}_i}
               \alpha_{i, \va_i}
                 = 1
           \right\rrbracket, \\
  B_j(\vv_j)
    &\defeq
           \left\llbracket
             \sum_{\vb_j \in \code{B}_j}
               \beta_{j, \vb_j} \vb_j
               = \vv_j
           \right\rrbracket \\
    &\quad\quad
           +
           \sum_{\vb_j \in \code{B}_j}
             \big\llbracket
               \beta_{j, \vb_j}
                 \geq 0
             \big\rrbracket
           +
           \left\llbracket
             \sum_{\vb_j \in \code{B}_j}
               \beta_{j, \vb_j}
                 = 1
           \right\rrbracket.
\end{align*}
With this, the global function of the FFG in
Fig.~\ref{fig:ffg:ldpc:code:binary:1:1} equals the augmented cost function
in~\eqref{eq:primal:lp:augmented:cost:function:1} and we have represented the
LP in terms of an FFG.\footnote{Note that instead of drawing function nodes
for the terms that appear in the definition of $A_i(\vu_i)$ and an edge for
the variables $\{ \alpha_{i,\va_i} \}_{\va_i \in \code{A}_i}$, we preferred to
simply draw a box for $A_i$, $i \in \set{I}$. A similar comment applies to
$B_j$, $j \in \set{J}$. An alternative approach would have been to apply the
concept of ``closing the box'' by Loeliger, cf.~e.g.~\cite{Loeliger:04:1},
where $A_i(\vu_i)$ would be defined as the minimum over $\{ \alpha_{i,\va_i}
\}_{\va_i \in \code{A}_i}$ of the above $A_i(\vu_i)$ function. Here we
preferred the first approach because we wanted to keep variables like $\vu_i$
and $\alpha_{i,\va_i}$ at the ``same
level''. \label{footnote:closing:the:box:comment:1}}

Of course, any reader who is familiar with LDPC codes will have no problem to
make a connection between the FFG of Fig. 1 and the standard representation as
a Tanner graph. Indeed, a node $A_i$ corresponds to a variable node in a
Tanner graph and a node $B_j$ takes over the role of a parity check node.
However, instead of simply assigning a variable to node $A_i$ we assign a
local set of constraints corresponding to the convex hull of a repetition
code. These are the equations $\sum_{\va_i\in {\cal{A}}_i} \alpha_{i,\va_i}
\va_i =\vu_i$, $\alpha_{i,\va_i}\geq 0$, $\sum_{\va_i\in {\cal{A}}_i}
\alpha_{i,\va_i} =1$. Similarly, the equations for the convex hull of a simple
parity-check code can be identified for nodes $B_j$.

\section{The Dual Linear Program}
\label{sec:dual:linear:program:1}

The dual linear program~\cite{Bertsimas:Tsitsiklis:97:1} of \textbf{PLPD2} is

\optprog
{
\begin{alignat*}{3}
  &
  \textbf{DLPD2:} \\
  &
  \text{max.} \quad
  &
    \sum_{i \in \set{I}}
      \phi'_i
  &
    +
    \sum_{j \in \set{J}}
      \theta'_j \\
  &
  \text{subj.~to } \quad
  &
    \phi'_i
    &\leq
       \min_{\va_i \in \code{A}_i}
         \langle
           -\vu'_i, \va_i
         \rangle
           \quad &&(i \in \set{I}), \\
  &&
  \theta'_j
    &\leq
       \min_{\vb_j \in \code{B}_j}
         \langle
           -\vv'_j, \vb_j 
         \rangle
           \quad &&(j \in \set{J}), \\
  &&
  u'_{i,j}
    &= - v'_{j,i}
         \quad &&((i,j) \in \set{E}), \\
  &&
  u'_{i,0}
    &= - x'_i 
         \quad &&(i \in \set{I}), \\
  &&
  x'_i
    &= \lambda_i
         \quad &&(i \in \set{I}).
\end{alignat*}
}{-0.8cm}

\noindent Expressing the constraints as additive cost terms, the above
maximization problem is equivalent to the (unconstrained) maximization of the
augmented cost function
\begin{align}
  &
  \sum_{i \in \set{I}}
    A'_i(\vu'_i)
  +
  \sum_{j \in \set{J}}
    B'_j(\vv'_j)
  -
  \sum_{(i,j) \in \set{E}}
    \llbracket
      u'_{i,j}
        = - v'_{j,i}
    \rrbracket
      \nonumber \\
  &
  -
  \sum_{i \in \set{I}}
    \llbracket
      u'_{i,0}
        = - x'_i
    \rrbracket
  -
  \sum_{i \in \set{I}}
    \llbracket
      x'_i
        = \lambda_i
    \rrbracket,
      \label{eq:dual:lp:augmented:cost:function:1}
\end{align}
with
\begin{align*}
  A'_i(\vu'_i)
    &= \phi'_i
       -
       \left\llbracket
         \phi'_i
           \leq 
             \min_{\va_i \in \code{A}_i}
             \langle
               -\vu'_i, \va_i 
             \rangle
       \right\rrbracket, \\
  B'_j(\vv'_j)
    &= \theta'_j
       -
       \left\llbracket
         \theta'_j
           \leq
             \min_{\vb_j \in \code{B}_j}
               \langle
                 -\vv'_j, \vb_j 
               \rangle
       \right\rrbracket.
\end{align*}
The augmented cost function in~\eqref{eq:dual:lp:augmented:cost:function:1} is
represented by the FFG in
Fig.~\ref{fig:dual:ffg:ldpc:code:binary:1:1}.\footnote{A similar comment
applies here as in Footnote~\ref{footnote:closing:the:box:comment:1}. Here,
the $\phi'_i$ and $\theta'_j$ have to be seen as dual variables that would
appear as edges in a more detailed drawing of the boxes $A'_i(\vu'_i)$ and
$B'_j(\vv'_j)$, respectively.} (For deriving \textbf{DLPD2} we used the
techniques introduced in~\cite{Vontobel:02:2, Vontobel:Loeliger:02:2}; note
that the techniques presented there can also be used to systematically derive
the dual function of much more complicated functions that are sums of convex
functions. Alternatively, one might also use results from monotropic
programming, cf.~e.g.~\cite{Bertsekas:99:1}.)

Because for each $i \in \set{I}$ the variable $\phi'_i$ is involved in only
one inequality, the optimal solution does not change if we replace the
corresponding inequality signs by equality signs in \textbf{DLPD2}. The same
comment holds for all $\theta'_j$, $j \in \set{J}$.

\begin{Definition}
  Let $\code{A} \defeq \code{A}_1 \times \cdots \times \code{A}_n$ and let
  $\code{B} \defeq \code{B}_1 \times \cdots \times \code{B}_m$.  For $\va
  \defeq (\va_1, \ldots, \va_n) \in \code{A}$ and $\vb \defeq (\vb_1,
  \ldots, \vb_m) \in \code{B}$ define
  \begin{align*}
    g'_{\va,\vb}(\vu')
      &\defeq 
         \sum_{i \in \set{I}}
           \langle
             -\vu'_i, \va_i
           \rangle
         +
         \sum_{j \in \set{J}}
           \langle
             +\vu'_j, \vb_j
           \rangle,
  \end{align*}
  where, with a slight abuse of notation, $\vu'_j$ is such that $[\vu'_j]_i =
  u'_{i,j}$ for all $(i,j) \in \set{E}$. Moreover, we call $(\va,\vb) \in
  \set{A} \times \set{B}$ consistent if $a_{i,j} = b_{j,i}$ for all $(i,j) \in
  \set{E}$. \edefinition
\end{Definition}

Obviously, $g'_{\va,\vb}(\vu')$ is a linear function in $\vu'$. With the above
definition, \textbf{DLPD2} can be rewritten to read

\optprog
{
\begin{alignat*}{3}
  &
  \textbf{DLPD3:} \\
  &
  \text{max.} \quad
  &
    g'_{\va,\vb}(\vu') \\
  &
  \text{subj.~to } \quad
  &
    (\va,\vb) \in \code{A} \times \code{B}, \\
  &&
    u'_{i,0} = -\lambda_i 
      && \quad (i \in \set{I}).
\end{alignat*}
}{-0.8cm}

\begin{Lemma}
  \label{lemma:consistent:assignment:1}

  Let $\vu'$ be such that $u'_{i,0} = -\lambda_i$, $i \in \set{I}$.  If
  $(\va,\vb) \in \set{A} \times \set{B}$ is consistent then
  $g'_{\va,\vb}(\vu')$ is constant in $\vu'$. Moreover, $g'_{\va,\vb}(\vu') =
  \langle \vlambda, \vx \rangle$, where $\vx$ is such that $x_i = a_{i,0}$, $i
  \in \set{I}$. If $(\va,\vb) \in \set{A} \times \set{B}$ is not consistent
  then $g'_{\va,\vb}(\vu')$ is not a constant function for at least one
  $u_{i,j}$, $(i,j) \in \set{E}$.
\end{Lemma}

\begin{Proof}
  See Sec.~\ref{sec:proof:lemma:consistent:assignment:1}.
\end{Proof}

\section{A Softened Dual Linear Program}
\label{sec:softened:dual:linear:program:1}

For any $\kappa \in \Rpp$, we define the soft-minimum operator to be
\begin{align*}
  {\min_{\ell}}^{(\kappa)}
    z_{\ell}
    &\defeq
       -
       \frac{1}{\kappa}
         \log
           \left(
             \sum_{\ell}
               \e^{-\kappa z_{\ell}}
           \right).
\end{align*}
(Note that $\kappa$ can be given the interpretation of an inverse
temperature.) One can easily check that ${\min_{\ell}}^{(\kappa)} z_{\ell}
\leq \min_{\ell} \{ z_{\ell} \}$ with equality in the limit $\kappa \to
+\infty$. Replacing the minimum operators in \textbf{DLPD2} by soft-minimum
operators, we obtain the modified optimization problem

\optprog
{
\begin{alignat*}{3}
  &
  \textbf{SDLPD2:} \\
  &
  \text{max.} \quad
  &
    \sum_{i \in \set{I}}
      \phi'_i
  &
    +
    \sum_{j \in \set{J}}
      \theta'_j \\
  &
  \text{subj.~to } \quad
  &
    \phi'_i
    &\leq
       {\min_{\va_i \in \code{A}_i}}^{(\kappa_i)}
         \langle
           -\vu'_i, \va_i
         \rangle
           \ &&(i \in \set{I}), \\
  &&
  \theta'_j
    &\leq
       {\min_{\vb_j \in \code{B}_j}}^{(\kappa_j)}
         \langle
           -\vv'_j, \vb_j 
         \rangle
           \ &&(j \in \set{J}), \\
  &&
  u'_{i,j}
    &= - v'_{j,i}
         \quad &&((i,j) \in \set{E}), \\
  &&
  u'_{i,0}
    &= - x'_i 
         \quad &&(i \in \set{I}), \\
  &&
  x'_i
    &= \lambda_i
         \quad &&(i \in \set{I}).
\end{alignat*}
}{-0.8cm}

\noindent In the following, unless noted otherwise, we will set $\kappa_i
\defeq \kappa$, $i \in \set{I}$, and $\kappa_j \defeq \kappa$, $j \in
\set{J}$, for some $\kappa \in \Rpp$. It is clear that in the limit $\kappa
\to +\infty$ we recover \textbf{DLPD2}.

\section{A Comment on the Dual of the Softened Dual Linear Program}
\label{sec:primal:softened:dual:linear:program:1}

Let
\begin{align*}
  H(\alpha_i)
    &\defeq 
       -
       \sum_{\va_i \in \code{A}_i}
         \alpha_{i,\va_i}
         \log(\alpha_{i,\va_i})
\end{align*}
be the entropy of of a random variable whose pmf takes on the values $\{
\alpha_{i,\va_i} \}_{\va_i \in \code{A}_i}$. Similarly, let
\begin{align*}
  H(\beta_j)
    &\defeq 
       -
       \sum_{\vb_j \in \code{B}_j}
         \beta_{j,\vb_j}
         \log(\beta_{j,\vb_j}).
\end{align*}
The dual of \textbf{SDLDP2} can then be written as

\optprog
{
\begin{alignat*}{2}
  &
  \textbf{DSLPD2:} \\
  &
  \text{min.} \quad
  &&
    \sum_{i \in \set{I}}
      \lambda_i x_i
    -
    \frac{1}{\kappa}
    \sum_{i \in \set{I}}
      H(\alpha_i)
    -
    \frac{1}{\kappa}
    \sum_{j \in \set{J}}
      H(\beta_j) \\
  &
  \text{subj.~to } \quad
  &&
  \text{the same constraints as in \textbf{PLPD2}}.
\end{alignat*}
}{-0.6cm}

\noindent We note that this is very close to the following Bethe free energy
optimization problem, cf.~e.g.~\cite{Yedidia:Freeman:Weiss:05:1}

\optprog
{
\begin{alignat*}{2}
  &
  \textbf{BFE1:} \\
  &
  \text{min.} \quad
  &&
    \sum_{i \in \set{I}}
      \lambda_i x_i \\
  &&&
    +
    \frac{1}{\kappa}
    \sum_{i \in \set{I}}
      (|\set{J}_i| - 1)
      H(\alpha_i)
    -
    \frac{1}{\kappa}
    \sum_{j \in \set{J}}
      H(\beta_j) \\
  &
  \text{subj.~to } \quad
  &&
  \text{the same constraints as in \textbf{PLPD2}},
\end{alignat*}
}{-0.6cm}

\noindent which, in turn,  can also be written as

\optprog
{
\begin{alignat*}{2}
  &
  \textbf{BFE2:} \\
  &
  \text{min.} \quad
  &&
    \sum_{i \in \set{I}}
      \lambda_i x_i
    -
    \frac{1}{\kappa}
    \sum_{i \in \set{I}}
      H(\alpha_i)
    -
    \sum_{j \in \set{J}}
      H(\beta_j) \\
  &&&
    +
    \frac{1}{\kappa}
    \sum_{i \in \set{I}}
      |\set{J}_i|
      H(\alpha_i)
     \\
  &
  \text{subj.~to } \quad
  &&
  \text{the same constraints as in \textbf{PLPD2}}.
\end{alignat*}
}{-0.6cm}

\noindent Without going into the details we note that the term
$+\frac{1}{\kappa} \sum_{i \in \set{I}} (|\set{J}_i|-1) H(\alpha_i)$ is
responsible for the fact that the cost function in \textbf{BFE2} is usually
non-convex for FFGs with cycles.

\section{Decoding Algorithm 1}
\label{sec:decoding:algorithm:1}

In the following, we assume that $u'_{i,j}$ and $v'_{j,i}$ are ``coupled'',
i.e.~we always have $u'_{i,j} = - v'_{j,i}$ for all $(i,j) \in \set{E}$.

The first algorithm that we propose is a coordinate-ascent-type algorithm for
solving \textbf{SDLPD2}. The main idea is to select edges $(i,j) \in \set{E}$
according to some update schedule: for each selected edge $(i,j) \in \set{E}$
we then replace the old values of $u'_{i,j}$, $\phi'_i$, and $\theta'_j$ by
new values such that the dual cost function is increased (or at least not
decreased). Practically, this means that we have to find an
$\overline{u}'_{i,j}$ such that $h'(\overline{u}'_{i,j}) \geq h'(u'_{i,j})$,
where
\begin{align*}
  h'(u'_{i,j})
    &\defeq
       {\min_{\va_i \in \code{A}_i}}^{(\kappa)}
         \langle
           -\vu'_i, \va_i
         \rangle
       +
       {\min_{\vb_j \in \code{B}_j}}^{(\kappa)}
         \langle
           -\vv'_j, \vb_j 
         \rangle.
\end{align*}
A simple way to achieve this is by setting
\begin{align}
  \overline{u}'_{i,j}
    &\defeq
       \arg \max_{u'_{i,j}}
         h'(u'_{i,j}).
           \label{eq:coordinate:ascent:edge:update:1:1}
\end{align}
The variables $\phi'_i$ and $\theta'_j$ are then updated accordingly so that
we obtain a new (dual) feasible point.

\begin{Lemma}
  \label{lemma:coordinate:ascent:update:rule:1}

  The value of $\overline{u}'_{i,j}$
  in~\eqref{eq:coordinate:ascent:edge:update:1:1} is given by
  \begin{align*}
    \overline{u}'_{i,j}
      &= \frac{1}{2}
           \bigg(
             +
             \big(
               S'_{i,0}
               -
               S'_{i,1}
             \big)
             -
             \big(
               T'_{j,0}
               -
               T'_{j,1}
             \big)
           \bigg),
  \end{align*}
  where
  \begin{align*}
    S'_{i,0}
      &\defeq
         -
         {\min_{\va_i \in \code{A}_i \atop a_{i,j} = 0}}^{(\kappa)}
           \langle
             -\tvu_i, \tva_i 
           \rangle, \\
    S'_{i,1}
      &\defeq
         -
         {\min_{\va_i \in \code{A}_i \atop a_{i,j} = 1}}^{(\kappa)}
           \langle
             -\tvu_i, \tva_i 
           \rangle, \\
    T'_{j,0}
      &\defeq
         -
         {\min_{\vb_j \in \code{B}_j \atop b_{j,i} = 0}}^{(\kappa)}
           \langle
             -
             \tvv_j, \tvb_j 
           \rangle, \\
    T'_{j,1}
      &\defeq
         -
         {\min_{\vb_j \in \code{B}_j \atop b_{j,i} = 1}}^{(\kappa)}
           \langle
             -\tvv_j, \tvb_j 
           \rangle.
  \end{align*}
  Here the vectors $\tvu$ and $\tva$ are the vectors $\vu$ and $\va$,
  respectively, where the $j$-th position has been omitted. Similarly, the
  vectors $\tvv$ and $\tvb$ are the vectors $\vv$ and $\vb$, respectively,
  where the $i$-th position has been omitted. Note that the differences
  $S'_{i,0} - S'_{i,1}$ and $T'_{i,0} - T'_{i,1}$, which are required for
  computing $\overline{u}'_{i,j}$, can be obtained very efficiently by using
  the sum-product algorithm~\cite{Kschischang:Frey:Loeliger:01}.
\end{Lemma}

\begin{Proof}
  See Sec.~\ref{sec:proof:lemma:coordinate:ascent:update:rule:1}.
\end{Proof}

In the introduction we wrote that we would like to use the special structure
of the primal/dual LP at hand;
Lemma~\ref{lemma:coordinate:ascent:update:rule:1} is a first example how this
can be done. Please note that when computing the necessary quantities (for the
case $\kappa = 1$) one has do computations that are (up to some flipped signs)
equivalent to computations that are done during message updates while
performing sum-product algorithm decoding of the LDPC code at hand.

\begin{Lemma}
  \label{lemma:coodinate:ascent:convergence:1}

  Assume that all the rows of the parity-check matrix $\matr{H}$ of the code
  $\code{C}$ have Hamming weight at least $3$.\footnote{Note that any
  interesting code has a parity-check matrix whose rows have Hamming weight at
  least $3$.} Then, updating cyclically all edges $(i,j) \in \set{E}$, the
  above coordinate-ascent algorithm converges to the maximum of
  \textbf{SDLPD2}.
\end{Lemma}

\begin{Proof}
  See Sec.~\ref{sec:proof:lemma:coodinate:ascent:convergence:1}
\end{Proof}

As we mentioned in the proof of
Lemma~\ref{lemma:coodinate:ascent:convergence:1}, the above algorithm can be
seen as a Gauss-Seidel-type algorithm. Let us remark that there are ways to
see sum-product algorithm decoding as applying a Gauss-Seidel-type algorithm
to the dual of the Bethe free energy, see e.g.~\cite{Tan:Rasmussen:05:1,
Walsh:Regalia:Johnson:05:1}; in light of the observations in
Sec.~\ref{sec:primal:softened:dual:linear:program:1} it is not surprising that
there is a tight relationship between our algorithms and the above-mentioned
algorithms.

\begin{Lemma}
  \label{lemma:coordinate:ascent:update:rule:2}

  For $\kappa \to \infty$, the function $h'(u'_{i,j})$ is maximized by any
  value $u'_{i,j}$ that lies in the closed interval between 
  \begin{align*}
    \big(
      S'_{i,0}
      -
      S'_{i,1}
    \big)
    \quad \text{ and } \quad
    -
    \big(
      T'_{j,0}
      -
      T'_{j,1}
    \big),
  \end{align*}
  where
  \begin{align*}
    S'_{i,0}
      &\defeq
         -
         \min_{\va_i \in \code{A}_i \atop a_{i,j} = 0}
           \langle
             -\tvu_i, \tva_i 
           \rangle, \\
    S'_{i,1}
      &\defeq
         -
         \min_{\va_i \in \code{A}_i \atop a_{i,j} = 1}
           \langle
             -\tvu_i, \tva_i 
           \rangle, \\
    T'_{j,0}
      &\defeq
         -
         \min_{\vb_j \in \code{B}_j \atop b_{j,i} = 0}
           \langle
             -
             \tvv_j, \tvb_j 
           \rangle, \\
    T'_{j,1}
      &\defeq
         -
         \min_{\vb_j \in \code{B}_j \atop b_{j,i} = 1}
           \langle
             -\tvv_j, \tvb_j 
           \rangle.
  \end{align*}
 
\end{Lemma}

\begin{Proof}
  See Sec.~\ref{sec:proof:lemma:coordinate:ascent:update:rule:2}.
\end{Proof}

\begin{Conjecture}
  \label{conj:coodinate:ascent:convergence:2}
 
  Again, we can cyclically update the edges $(i,j) \in \set{E}$ whereby the
  new $\overline{u}'_{i,j}$ is chosen randomly in the above interval. Although
  the objective function for $\kappa \to +\infty$ is concave, it is not
  everywhere differentiable. This makes a convergence proof in the style of
  Lemma~\ref{lemma:coodinate:ascent:convergence:1} difficult. We think that we
  can again use the special structure of the LP at hand to show that the
  algorithm cannot get stuck at a suboptimal point. However, so far we do not
  have a proof of this
  fact. Sec.~\ref{sec:considerations:conj:coodinate:ascent:convergence:2}
  discusses briefly why a convergence proof is not a trivial extension of
  Lemma~\ref{lemma:coodinate:ascent:convergence:1}. \econjecture
\end{Conjecture}

Before ending this section, let us briefly remark how a codeword decision is
obtained from a solution of \textbf{DLPD2}. Assume that $\hat \vx$ is the
pseudo-codeword that is the solution to \textbf{PLPD1} or to
\textbf{PLPD2}.\footnote{We assume here that there is a unique optimal
solution $\hat \vx$ to \textbf{PLPD1} or to \textbf{PLPD2}; more general
statements can be made for the case when there is not a unique optimal
solution.}  Knowing the solution of \textbf{DLPD2} we cannot directly find
$\hat \vx$, however, we can find out at what positions $\hat \vx$ is $0$ and
at what positions $\hat \vx$ is $1$. Namely, letting $\check \vx \in \{ 0, ?,
1 \}^n$ have the components
\begin{align*}
  \check x_i
    \defeq
      \begin{cases}
        0  & \text{ if }
             \langle
               - \vu'_i, \va_i
	     \rangle
	     |_{\va_i = \vect{0}}
             \ < \ 
	     \langle
	       - \vu'_i, \va_i
             \rangle
             |_{\va_i = \vect{1}} \\
        ?  & \text{ if }
             \langle
               - \vu'_i, \va_i
	     \rangle
	     |_{\va_i = \vect{0}}
	     \ = \ 
	     \langle
	       - \vu'_i, \va_i
	     \rangle
	     |_{\va_i = \vect{1}} \\
        1  & \text{ if }
             \langle
	       - \vu'_i, \va_i
	     \rangle
	     |_{\va_i = \vect{0}}
	     \ > \ 
	     \langle
	       - \vu'_i, \va_i
	     \rangle
	     |_{\va_i = \vect{1}}
      \end{cases},
\end{align*}
we have $\check x_i = \hat x_i$ when $\hat x_i$ equals $0$ or $1$ and $\check
x_i = \ ?$ when $\hat x_i \in (0,1)$. In other words, with the solution to
\textbf{DLPD2} we do not get the exact $\hat \vx$ in case $\hat \vx$ is not a
codeword. However, as a side remark, because $\supp(\hat \vx) = \supp(\check
\vx)$ (where $\supp$ is the set of all non-zero positions) we can use $\check
\vx$ to find the stopping
set~\cite{Di:Proietti:Telatar:Richardson:Urbanke:02:1} associated to $\hat
\vx$.

\section{Decoding Algorithm 2}
\label{sec:decoding:algorithm:2}

Again, we assume that $u'_{i,j}$ and $v'_{j,i}$ are ``coupled'', i.e.~we
always have $u'_{j,i} = - v'_{i,j}$ for all $(i,j) \in \set{E}$.

While the iterative solutions of the coordinate-ascent methods that we
presented in the previous section resemble the traditional min-sum algorithm
decoding rules (and sum-product algorithm decoding rules) relatively closely,
other methods for solving the linear program also offer attractive
complexity/performance trade-offs. We would like to point out one such
algorithm which is well suited for the linear programming problem arising from
the decoding setup. Indeed, observing the formulation of the dual linear
program \textbf{DLPD2}, sub-gradient methods\footnote{The use of sub-gradients
is necessary since the objective function is concave but not everywhere
differentiable, cf.~e.g.~\cite{Bertsekas:99:1}.} are readily available to
perform the required maximization. However, in order to exploit the structure
of the problem we focus our attention to incremental sub-gradient
methods~\cite{Nedic:02:1}. Algorithms belonging to this family of optimization
procedures allow us to exploit the fact that the objective function is a sum
of a number of terms and we can operate on each term, i.e.~each constituent
code in the FFG, individually. In order to derive a concise formulation of the
procedure we start by considering a check node $j \in \set{J}$. For a
particular choice of dual variables $\vv'_j$ the contribution of node $j$ to
the overall objective function is
\begin{align*}
  f'_j(\vv'_j)
    &= \min_{\vb_j \in \code{B}_j}
         \langle
           -\vv'_j, \vb_j
         \rangle.
\end{align*}
Let a function $\vg'(\vv'_j)$ be defined as $\vg'_j(\vv'_j) \defeq - \arg
\min_{\vb_j \in \code{B}_j} \langle -\vv'_j, \vb_j \rangle$ where, if
ambiguities exist, $\vg'_j(\vv'_j)$ is the negative of an arbitrary
combination of the set of ambiguous vectors $\vb'_j$. Note that for obtaining
$\vg'_j(\vv'_j)$ we can again take advantage of the special structure of the
LP at hand.

Using the defining property of sub-gradient $\vd'_j$ at $\vv'_j$, namely,
\begin{align*}
  f'(\uvv'_j)
    &\leq f(\vv'_j)
          + 
          \langle
            \vd'_j, 
            \uvv'_j - \vv'_j)
\end{align*}
it can be seen that $\vg'_j(\vv'_j)$ is a sub-gradient. We can then update
$\vv'_j$ as follows:
\begin{align*}
  \vv'_j 
    &\leftarrow
       \vv'_j
       +
       \mu_{\ell} \vg'_j(\vv'_j),
\end{align*}
where $\mu_{\ell} \in \Rpp$. Given this, one can formulate the
following algorithm: at iteration $\ell$ update consecutively all check nodes
$j \in \set{J}$ and then, in an analogous manner, update all variable nodes $i
\in \set{I}$.

For this algorithm we cannot guarantee that the value of the objective
function increases for each iteration (not even for small
$\mu_{\ell}$). Nevertheless, its convergence to the maximum can be
guaranteed for a suitably chosen sequence $\{ \alpha_\ell \}_{\ell \geq
1}$~\cite{Nedic:02:1}.

Let us point out that gradient-type methods have also been used to decode
codes in different contexts, see e.g.~the work by Lucas et
al.~\cite{Lucas:Bossert:Breitbach:98:1}. However, the setup
in~\cite{Lucas:Bossert:Breitbach:98:1} has some significant differences to the
setup here: firstly, the objective function of the optimization problem
in~\cite{Lucas:Bossert:Breitbach:98:1} does not depend on the observed
log-liklihood ratio vector $\vlambda$, secondly, the starting point
in~\cite{Lucas:Bossert:Breitbach:98:1} is chosen as a function of $\vlambda$.

\section{Simulation Results}
\label{sec:simulation:results:1}

\begin{figure}[t]
  \begin{center}
    \epsfig{file=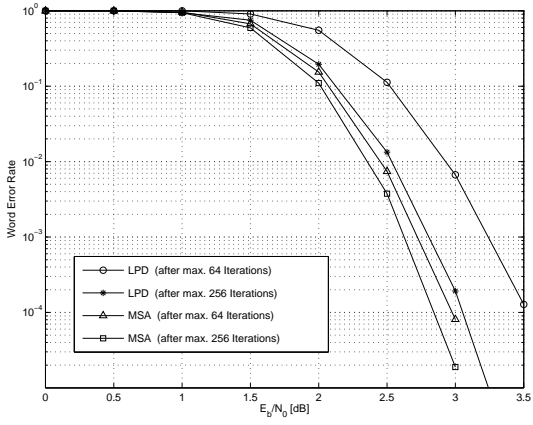, width=0.90\linewidth}
  \end{center}
  \caption{Decoding results for a $[1000,500]$ LDPC code. (See
           Sec.~\ref{sec:simulation:results:1} for more details.)}
  \label{fig:simulation:results:1:1}
\end{figure}

As a proof of concept we show some simulation results for a randomly generated
$(3,6)$-regular $[1000,500]$ LDPC code where four-cycles in the Tanner graph
have been eliminated. Fig.~\ref{fig:simulation:results:1:1} shows the decoding
results based on Decoding Algorithm $1$ with update rule
Lemma~\ref{lemma:coordinate:ascent:update:rule:2} compared with standard
min-sum algorithm decoding~\cite{Kschischang:Frey:Loeliger:01}.

\section{Conclusions}
\label{sec:conclusions:1}

We have discussed some initial steps towards algorithms that are specially
targeted for efficiently solving the LP that appears in LP decoding. It has
been shown that algorithms with memory and time complexity similar to min-sum
algorithm decoding can be achieved. There are many avenues to pursue this
topic further, e.g.~by improving the update schedule, by studying how to
design codes that allow efficient hardware implementation of the proposed
algorithms, or by investigating other algorithms that use the structure of the
LP that appears in LP decoding. We hope that this paper raises the interest in
exploring these research directions.

Finally, without going into the details, let us remark that the algorithms
here can also be used to solve certain linear programs whose value can be used
to obtain lower bounds on the minimal AWGNC pseudo-weight of parity-check
matrices, cf.~\cite[Claim 3]{Vontobel:Koetter:04:1}. (Actually, one does not
really need to solve the linear program
in~\cite[Claim~3]{Vontobel:Koetter:04:1} in order to obtain a lower bound on
the minimum AWGNC pseudo-weight, any dual feasible point is good enough for
that purpose.)

\appendix

\subsection{Proof of Lemma~\ref{lemma:consistent:assignment:1}}
\label{sec:proof:lemma:consistent:assignment:1}

If $(\va,\vb)$ is consistent then 
\begin{align*}
  g'_{\va,\vb}(\vu')
    &= \sum_{i \in \set{I}}
         \langle
           -\vu'_i, \va_i
         \rangle
       +
       \sum_{j \in \set{J}}
         \langle
           +\vu'_j, \vb_j
         \rangle \\
    &= -
       \sum_{i \in \set{I}}
         u'_{i,0} a_{i,0}
       -
       \sum_{(i,j) \in \set{E}}
         u'_{i,j} a_{i,j} \\
    &\quad\,
       +
       \sum_{(i,j) \in \set{E}}
         u'_{i,j} b_{j,i} \\
    &= -
       \sum_{i \in \set{I}}
         u'_{i,0} a_{i,0} \\
    &= \sum_{i \in \set{I}}
         \lambda_i x_i.
\end{align*}
On the other hand, if $(\va,\vb)$ is not consistent and $(i,j) \in \set{E}$ is
such that $a_{i,j} \neq b_{j,i}$ then $g'_{\va,\vb}(\vu')$
is non-constant in $u'_{i,j}$.

\subsection{Proof of Lemma~\ref{lemma:coordinate:ascent:update:rule:1}}
\label{sec:proof:lemma:coordinate:ascent:update:rule:1}

This result is obtained by taking the derivative of $h'(u'_{i,j})$, setting it
equal to zero, and solving for $u'_{i,j}$. Let us go through this procedure
step by step. Using the fact that $u'_{i,j} = -v'_{j,i}$, the function
$h'(u'_{i,j})$ can be written as
\begin{align*}
  h'(u'_{i,j})
    &\defeq
       {\min_{\va_i \in \code{A}_i}}^{(\kappa)}
         \langle
           -\vu'_i, \va_i
         \rangle
       +
       {\min_{\vb_j \in \code{B}_j}}^{(\kappa)}
         \langle
           -\vv'_j, \vb_j 
         \rangle \\
  &= -
     \frac{1}{\kappa}
     \log
       \left(
         \sum_{\va_i \in \code{A}_i}
         \e^
         {
           +
           \kappa
           \langle
             \vu_i, \va_i 
           \rangle
         }
       \right) \\
  &\quad\,
     -
     \frac{1}{\kappa}
     \log
       \left(
         \sum_{\vb_j \in \code{B}_j}
         \e^
         {
           +
           \kappa
           \langle
             \vv_j, \vb_j 
           \rangle
         }
       \right) \\
  &= -
     \frac{1}{\kappa}
     \log
       \left(
         \sum_{\va_i \in \code{A}_i}
         \e^
         {
           +
           \kappa u_{i,j} a_{i,j} 
           +
           \kappa
           \langle
             \tilde \vu_i, \tva_i 
           \rangle
          }
       \right) \\
     &\quad\,
     -
     \frac{1}{\kappa}
     \log
       \left(
         \sum_{\vb_j \in \code{B}_j}
         \e^
         {
           -
           \kappa u_{i,j} b_{j,i} 
           +
           \kappa
           \langle
             \tilde \vv_j, \tvb_j 
           \rangle
          }
       \right) \\
     &= -
       \frac{1}{\kappa}
       \log
         \left(
           \e^{\kappa S'_{i,0}}
           +
           \e^{+\kappa u'_{i,j}}
           \e^{\kappa S'_{i,1}}
         \right) \\
     &\quad\,
       -
       \frac{1}{\kappa}
       \log
         \left(
           \e^{\kappa T'_{j,0}}
           +
           \e^{-\kappa u'_{i,j}}
           \e^{\kappa T'_{j,1}}
         \right)
\end{align*}
Setting the derivative of $h'(u'_{i,j})$ with respect to $u'_{i,j}$ equal to
zero we obtain
\begin{align*}
  0
    &\overset{!}{=}
       \frac{\partial h'(u'_{i,j})}
            {\partial u'_{i,j}} \\
  &= -
     \frac{1}{\kappa}
     \cdot
     \frac{+\kappa \e^{+\kappa u'_{i,j}} \e^{\kappa S'_{i,1}}}
          {\e^{\kappa S'_{i,0}}
           +
           \e^{+\kappa u'_{i,j}}
           \e^{\kappa S'_{i,1}}
          } \\
  &\quad\,
     -
     \frac{1}{\kappa}
     \cdot
     \frac{-\kappa \e^{-\kappa u'_{i,j}} \e^{\kappa T'_{j,1}}}
          {
           \e^{\kappa T'_{j,0}}
           +
           \e^{-\kappa u'_{i,j}}
           \e^{\kappa T'_{j,1}}
          }.
\end{align*}
Multiplying out we get
\begin{align*}
  &
  \e^{+\kappa u'_{i,j}} \e^{\kappa (S'_{i,1} + T'_{j,0})}
  +
  \e^{\kappa (S'_{i,1} + T'_{j,1})} \\
  &\quad
   = \e^{-\kappa u'_{i,j}} \e^{\kappa (S'_{i,0} + T'_{j,1})} 
     +
     \e^{\kappa (S'_{i,1} + T'_{j,1})}.
\end{align*}
This yields
\begin{align*}
  \overline{u}'_{i,j}
    &= \frac{1}{2}
         \bigg(
           +
           \big(
             S'_{i,0}
             -
             S'_{i,1}
           \big)
           -
           \big(
             T'_{j,0}
             -
             T'_{j,1}
           \big)
         \bigg),
\end{align*}
which is the promised result.

\subsection{Proof of Lemma~\ref{lemma:coodinate:ascent:convergence:1}}
\label{sec:proof:lemma:coodinate:ascent:convergence:1}

We can use results from~\cite[Sec.~2.7]{Bertsekas:99:1}, where the following
setup is considered.\footnote{We have adapted the text for maximizations
instead of minimizations.} Consider the optimization problem

\optprog
{
\begin{alignat*}{2}
  &
  \text{maximize} \quad
  &&
    f(\vx) \\
  &
  \text{subject to } \quad
  &&
   \vx \in \set{X},
\end{alignat*}
}{-0.8cm}

\noindent where $\set{X} \defeq \set{X}_1 \times \cdots \times \set{X}_m$. The
set $\set{X}_i$ is assumed to be a closed convex subset of $\R^{n_i}$ and $n =
n_1 + \cdots + n_m$. The vector $\vx$ is partitioned as $\vx = (\vx_1, \ldots,
\vx_m)$ where each $\vx_i \in \R^{n_i}$. So the constraint $\vx \in \set{X}$
is equivalent to $\vx_i \in \set{X}_i$, $i \in \{ 1, \ldots, m \}$.

The following algorithm, known as block coordinate-ascent or non-linear
Gauss-Seidel method, generates the next iterate $\vx^{k+1} \defeq
(\vx_1^{k+1}, \ldots, \vx_m^{k+1})$, given the current iterate $\vx^k \defeq
(\vx_1^k, \ldots, \vx_m^k)$ according to the iteration
\begin{align}
  \vx_i^{k+1}
    &\defeq
       \arg \max_{\vxi_i \in \set{X}_i} 
         \nonumber \\
    &\quad\quad
         f(\vx_1^{k+1}, \ldots, \vx_{i-1}^{k+1}, 
           \vxi_i, 
           \vx_{i+1}^k, \ldots, \vx_m^k).
             \label{eq:bertsekas:2}
\end{align}

\begin{Proposition}[{\cite[Prop.~2.7.1]{Bertsekas:99:1}}]
  Suppose that $f$ is continuously differentiable over the set
  $\set{X}$. Furthermore, suppose that for each $i$ and $\vx \in \set{X}$, the
  maximum below
  \begin{align*}
    \max_{\vxi_i \in \set{X}_i}
      f(\vx_1, \ldots, \vx_{i-1}, \vxi_i, \vx_{i+1}, \ldots, \vx_m)
  \end{align*}
  is uniquely attained. Let $\{ \vx^k \}$ be the sequence generated by the
  block coordinate-ascent method~\eqref{eq:bertsekas:2}. Then every limit
  point of $\{ \vx^k \}$ is a stationary point. \eproposition
\end{Proposition}

We turn our attention now to our optimization problem. The fundamental
polytope (which is the set $\bigcap_{j \in \set{J}} \convhull(\code{C}_j)$),
has dimension $n$ if and only if the parity-check matrix has no rows of
Hamming weight $1$ and $2$. This type of non-degeneracy of \textbf{PLPD2}
implies the strict concavity of the function that we try to optimize in
\textbf{SDLPD2}. Based on $\vlambda$ one can then without loss of generality
define suitable closed intervals for each variable so that one can apply the
above proposition to our algorithm.

\subsection{Proof of Lemma~\ref{lemma:coordinate:ascent:update:rule:2}}
\label{sec:proof:lemma:coordinate:ascent:update:rule:2}

\begin{figure}[t]
  \begin{center}
    \epsfig{file=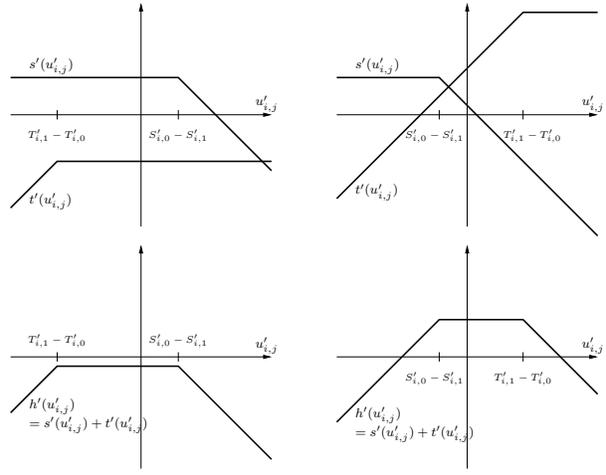, width=1.0\linewidth}
  \end{center}
  \caption{Illustration of the functions $s'(u'_{i,j})$, $t'(u'_{i,j})$, and
    $h'(u'_{i,j})$ appearing the the proof of
    Lemma~\ref{lemma:coordinate:ascent:update:rule:2}. Left plots: exemplary
    case for $S'_{i,0} - S'_{i,1} \geq T'_{i,1} - T'_{i,0}$. Right plots:
    exemplary case for $S'_{i,0} - S'_{i,1} \leq T'_{i,1} - T'_{i,0}$.}
  \label{fig:critical:points:1:1}
\end{figure}

Define the functions
\begin{align*}
  s'(u'_{i,j})
    &\defeq
       \min_{\va_i \in \code{A}_i}
         \langle
           -\vu'_i, \va_i
         \rangle
           \quad \text{ and } \\
  t'(u'_{i,j})
    &\defeq
       \min_{\vb_j \in \code{B}_j}
         \langle
           -\vv'_j, \vb_j 
         \rangle
\end{align*}
such that $h'(u'_{i,j}) = s'(u'_{i,j}) + t'(u'_{i,j})$. Then
\begin{align*}
  s'(u'_{i,j})
    &\defeq
       \min_{\va_i \in \code{A}_i}
         \langle
           -\vu'_i, \va_i
         \rangle \\
    &= \min_{\va_i \in \code{A}_i}
         -
         u'_{i,j} a_{i,j} 
         -
         \langle
           \tilde \vu'_i, \tva_i 
         \rangle \\
    &= \min
         \big(
           -S'_{i,0}, \ - u'_{i,j} {-} S'_{i,1}
         \big), \\
  t'(u'_{i,j})
    &\defeq
       \min_{\vb_j \in \code{B}_j}
         \langle
           -\vv'_j, \vb_j 
         \rangle \\
    &= \min_{\vb_j \in \code{B}_j}
         +
         u'_{i,j} b_{j,i} 
         -
         \langle
           \tilde \vv'_j, \tvb_j 
         \rangle \\
    &= \min
         \big(
           -T'_{i,0}, \ + u'_{i,j} {-} T'_{i,1}
         \big).
\end{align*}
As can be seen from Fig.~\ref{fig:critical:points:1:1}, the functions
$s'(u'_{i,j})$ and $t'(u'_{i,j})$ are both piece-wise linear
functions. Whereas the function $s'(u'_{i,j})$ is flat up to $u'_{i,j} =
S'_{i,0} - S'_{i,1}$ and then has slope $-1$, the function $t'(u'_{i,j})$
increases with slope $+1$ up to $u'_{i,j} = T'_{i,1} - T'_{i,0}$ and is then
flat. From Fig.~\ref{fig:critical:points:1:1} is can also be seen that,
independently if $S'_{i,0} - S'_{i,1}$ is larger or smaller than $T'_{i,1} -
T'_{i,0}$, the function $h'(u'_{i,j})$ always consists of three parts: first
it increases with slope $+1$, then it is flat, and finally it decreases with
slope $-1$. From this observations, the lemma statement follows.

\subsection{Comment to Conjecture~\ref{conj:coodinate:ascent:convergence:2}}
\label{sec:considerations:conj:coodinate:ascent:convergence:2}

\begin{figure}[t]
  \begin{center}
    \epsfig{file=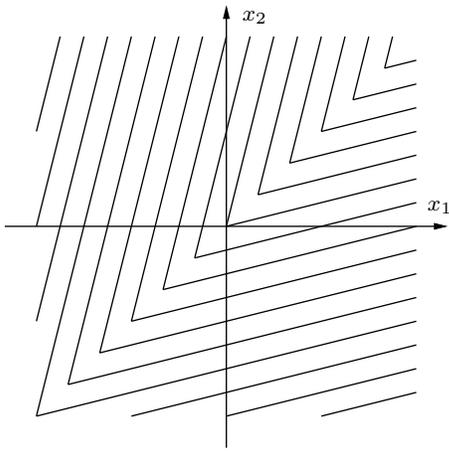, height=0.75\linewidth}
  \end{center}
  \caption{Level curves of $f(x_1, x_2)$ in
    Sec.~\ref{sec:considerations:conj:coodinate:ascent:convergence:2}.}
  \label{fig:problematic:contour:1:1}
\end{figure}

This section briefly discusses a concave function where a coordinate-ascent
approach does not find the global maximum. Let $0 < a < 1$ and let
\begin{align*}
  f(x_1, x_2)
     \defeq
       \min \big(\!
              & - x_1 + x_2 + a (x_1 + x_2), \\
              & + x_1 - x_2 + a (x_1 + x_2)
            \big).
\end{align*}
The level curves of $f(x_1, x_2)$ are shown in
Fig.~\ref{fig:problematic:contour:1:1}. By choosing $(x_1, x_2) \defeq
(\alpha, \alpha)$ and letting $\alpha$ go to $\infty$ we see that this
function is unbounded.

Consider now the optimization problem

\optprog
{
\begin{alignat*}{2}
  \quad
  &
  \text{maximize} \quad
  &&
    f(x_1, x_2) \\
  &
  \text{subject to } \quad
  &&
   (x_1, x_2) \in \set{X},
\end{alignat*}
}{-0.8cm}

\noindent where $\set{X}$ is some suitably chosen closed convex subset of
$\R^2$. Assume that a coordinate-ascent-type method has e.g. found the point
$(x_1, x_2) = (0,0)$ with $f(0,0) = 0$. (Of course, we assume that $(0,0) \in
\set{X}$.) Unfortunately, at this point the coordinate-ascent-type method
cannot make any progress because $f(x_1,0) = \min \big(-(1-a)x_1, (1+a) x_1
\big) < 0$ for all $x_1 \neq 0$ and $f(0,x_2) = \min \big((1+a)x_2, -(1-a)x_2
\big) < 0$ for all $x_2 \neq 0$.

However, defining 
\begin{align*}
  f^{(\kappa)}(x_1, x_2)
     \defeq
       {\min}^{(\kappa)}
            \big(\!
              & - x_1 + x_2 + a (x_1 + x_2), \\
              & + x_1 - x_2 + a (x_1 + x_2)
            \big),
\end{align*}
where $\kappa \in \Rpp$ is arbitrary, a coordinate-ascent method can
successfully be used for the ``softened'' optimization problem

\optprog
{
\begin{alignat*}{2}
  \quad
  &
  \text{maximize} \quad
  &&
    f^{(\kappa)}(x_1, x_2) \\
  &
  \text{subject to } \quad
  &&
   (x_1, x_2) \in \set{X}.
\end{alignat*}
}{-0.8cm}

\end{document}